\documentclass[twocolumn,showpacs,preprintnumbers,amsmath,amssymb]{revtex4}
\usepackage{amssymb}
\usepackage{color}
\usepackage{graphicx}% Include figure files
\usepackage{dcolumn}% Align table columns on decimal point
\usepackage{bm}% bold math
\usepackage{array}
\usepackage[colorlinks,
citecolor=blue]{hyperref}
\newcommand{\PreserveBackslash}[1]{\let\temp=\\#1\let\\=\temp}
\newcolumntype{C}[1]{>{\PreserveBackslash\centering}p{#1}}
\newcolumntype{R}[1]{>{\PreserveBackslash\raggedleft}p{#1}}
\newcolumntype{L}[1]{>{\PreserveBackslash\raggedright}p{#1}}

\begin{document}

\title{Evaluating Network Models: A Likelihood Analysis}

\author{Wen-Qiang Wang$^1$}
\author{Qian-Ming Zhang$^{1,2}$}
\author{Tao Zhou$^1$}
\email{zhutou@ustc.edu} \affiliation{$^1$Web Sciences Center,
School of Computer Science and Technology, University of
Electronic Science and Technology of China, 610054 Chengdu, People's Republic of China\\
$^2$Beijing Computational Science Research Center, Beijing 100089,
People's Republic of China}

\date{\today}

\begin{abstract}
Many models are put forward to mimic the evolution of real
networked systems. A well-accepted way to judge the validity is to
compare the modeling results with real networks subject to several
structural features. Even for a specific real network, we cannot
fairly evaluate the goodness of different models since there are
too many structural features while there is no criterion to select
and assign weights on them. Motivated by the studies on link
prediction algorithms, we propose a unified method to evaluate the
network models via the comparison of the likelihoods of the
currently observed network driven by different models, with an
assumption that the higher the likelihood is, the better the model
is. We test our method on the real Internet at the Autonomous
System (AS) level, and the results suggest that the Generalized
Linear Preferential (GLP) model outperforms the Tel Aviv Network
Generator (Tang), while both two models are better than the
Barab\'asi-Albert (BA) and Erd\"os-R\'enyi (ER) models. Our method
can be further applied in determining the optimal values of
parameters that correspond to the maximal likelihood. Experiment
indicates that the parameters obtained by our method can better
capture the characters of newly-added nodes and links in the
AS-level Internet than the original methods in the literature.
\end{abstract}


\maketitle

\section{Introduction}

Recent years have witnessed a fast development of complex networks
\cite{Albert_rmp_2002, Dorogovtsev_aip_2002, Newman_siam_2003,
Barabasi_science_2009}. A network is a set of items that are
called vertices with connections between them, which are named as
edges. Many natural and man-made systems can be described as
networks. Such paragons cannot be numbered that biological
networks including protein-protein interaction networks
\cite{HJ_nature_2001} and metabolic network \cite{HJ_nature_2000};
social networks such as movie actor collaboration
\cite{He_physicaa_2006} and scientific collaboration networks
\cite{Newman_pnas_2001}; technological networks like power grids
\cite{Watts_nature_1998}, WWW \cite{Albert_nature_1999} and the
Internet at the Autonomous System (AS) level
\cite{Bu_infocom_2002, Zhou_2003, Park_infocom_2004,
Zhou_pre_2004, Bar_lncs_2004, Zhang_njp_2008}. A major endeavor in
academics is to discover the common properties shared by many real
networks and the specific features owned by a certain type of
networks. A great number of measurements to reveal the structural
features of networks are applied \cite{Costa_ap_2007}. The degree
distribution \cite{Clauset_siam_2009}, as one of the most
important global measurements, has attracted increasing attention
since the awareness of the scale-freeness
\cite{Barabasi_science_1999}. Clustering coefficient is a local
measurement that characterizes the loop structure of order three.
Another significant measurement is the average distance. A network
is considered to be small-world if it has large clustering
coefficient but short average distance \cite{Watts_nature_1998}.
Except for the properties mentioned above, there are many other
measurements such as degree-degree correlation
\cite{Callaway_pre_2001}, betweenness centrality
\cite{Freeman_sociometry_1977} and so forth. Moreover, some
statistical measurements borrowed from physics such as entropy
\cite{Shannon_BSTJ_1948}, and novel metrics such as modularity
\cite{Newman_pnas_2006} also play important roles in
characterizing networks.

Not only the statistical features but also the dynamical evolution
of networks the current research interest has focused on. A mess
of models have been proposed to reveal the origins of the
impressive statistical features of complex networks. There are
also many evolving models developed for some certain type of
networks such as the Internet at the AS level
\cite{Bu_infocom_2002, Zhou_2003, Park_infocom_2004,
Zhou_pre_2004, Bar_lncs_2004, Zhang_njp_2008}, the social networks
\cite{Viscek_physciaa_2002, Arenas_pre_2004, Kumar_2010,
Huang_wsdm_2008, Albert_prl_2000, Dorogovtsev_pre_2000} and so
forth. However the prosperous development of measurements sets a
barrier for evaluating different evolving models. The traditional
idea is that: if the network generated by a model resembles the
target network in terms of some statistical features usually
selected by the authors themselves, the model is claimed as a
proper description of the real evolution. But this methodology
seems to be puzzling. First, unselected statistical properties are
entirely ignored so no one knows whether the model is sufficient
to describe them as well. Secondly, the authors tend to select the
metrics that support their models. Therefore, it is impossible to
give a fair remark that which model is better. Thirdly, it is
difficult to quantify the extent to which the models resemble the
real evolving mechanisms.

Inspired by the link prediction approaches and likelihood
analysis, we propose a method that tries to fairly and objectively
evaluate different models. Link prediction aims at estimating the
likelihood of non-existing edges in a network and try to dig out
the missing edges \cite{Lv_physicaa_2011}. The evolution of
networks involves two processes - one is the addition or deletion
of nodes and another one is the changing of edges between nodes
\cite{Albert_prl_2000}. In principle the rules of the additions of
edges of a model can be considered as a kind of link prediction
algorithm and here lies the bridge between link prediction and the
mechanism of evolving models.

The present paper is organized as follows. We will give a general
description of our method in Section II. Section III introduces
the data and explains how to use our method to evaluate evolving
models in details with the AS-level Internet being an example
network. The results obtained by our method are shown in Section
IV. We draw the conclusion and give some discussion in the last
section.

\section{Method}

In this section, we will give a general description about our
method to evaluate evolving models. It is believed that an
evolving model is a description of the evolving process of a
network in reality. An evolving model describes the evolving
mechanism of a real network or a class of networks. Given two
snaps of one network at time $t_{1}$ and $t_{2}$ ($t_{1}<t_{2}$),
as well as an evolving model, we can in principle calculate the
likelihood that the network starting from the configuration at
time $t_{1}$ will evolves to the configuration at $t_{2}$ under
the rules of the given model. We say a model is \emph{better} than
another one if the likelihood of the former model is greater than
that of the latter one. However, how to calculate such likelihood
is still a big challenge. Inspired by the like prediction
algorithms, we can calculate the likelihood of the addition of an
edge according to a given evolving model \cite{Lv_physicaa_2011}.
In a short duration of time, each edge's generation can be thought
as independent to others and the sequence of generations can be
ignored. Thus the likelihood mentioned above is the product of the
newly generated edges' likelihoods.

Denote by $G$ the network and $E_{t}$ the set of edges at time
step $t$. The new edges generated at the current time step is
$E_{new}=E_{t+1} \backslash E_{t}$. The probability that node $i$
is selected as one end of the newly generated edge is
\begin{equation}\label{eq:a}
\Pi_{i}=f(G, \vec a),
\end{equation}
where $\vec a$ is the set of parameters applied by the model. Then the likelihood of a new monitored edge $(i,j)$ is
\begin{equation}\label{eq:b}
P_{(i,j)}=\Pi_{i} \times \Pi_{j}.
\end{equation}
Eq. (\ref{eq:b}) is applicable only when $i$ and $j$ are both old
nodes. If $i$ or $j$ is newly generated, we set $\Pi_{i}=1$ or
$\Pi_{j}=1$. In order to make comparison between different models,
$P(i,j)$ is normalized by $1/\sum_{(a,b) \in E^{N}}{p(a,b)}$,
where $E^{N}$ is the set of nonexisting edges($(i,j) \in E^{N}$).
Given different parameters $\vec{a}$, the values of $P(i,j)$ may
be different, resulting in different likelihoods of the target
network. The parameters corresponding to the maximum likelihood
are intuitively considered to be the optimal set of parameters for
the evaluated model. In a word, a network's likelihood can be
calculated if the evolution data and the corresponding model are
given. And if there are several candidate models, our method could
judge them by comparing the corresponding likelihoods: the model
giving higher likelihood according to the target network is more
favored.
\begin{figure*}[htb]\makeatletter\def\@captype{figure}\makeatother
\includegraphics[width=0.9\textwidth]{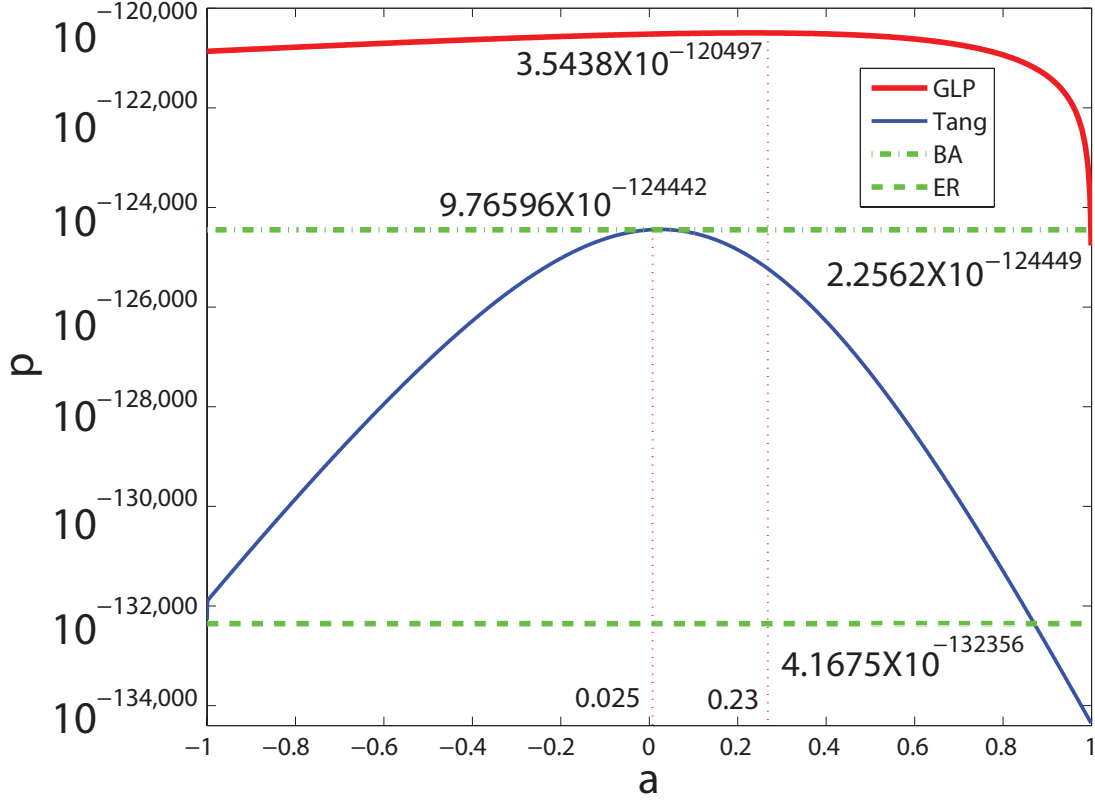}
\caption{Likelihoods for different models and different parameters.}.
\label{fig:figure1}
\end{figure*}

\section{Experimental Analysis}

\begin{figure*}[htb]\makeatletter\def\@captype{figure}\makeatother
\centering
\includegraphics[width=0.9\textwidth]{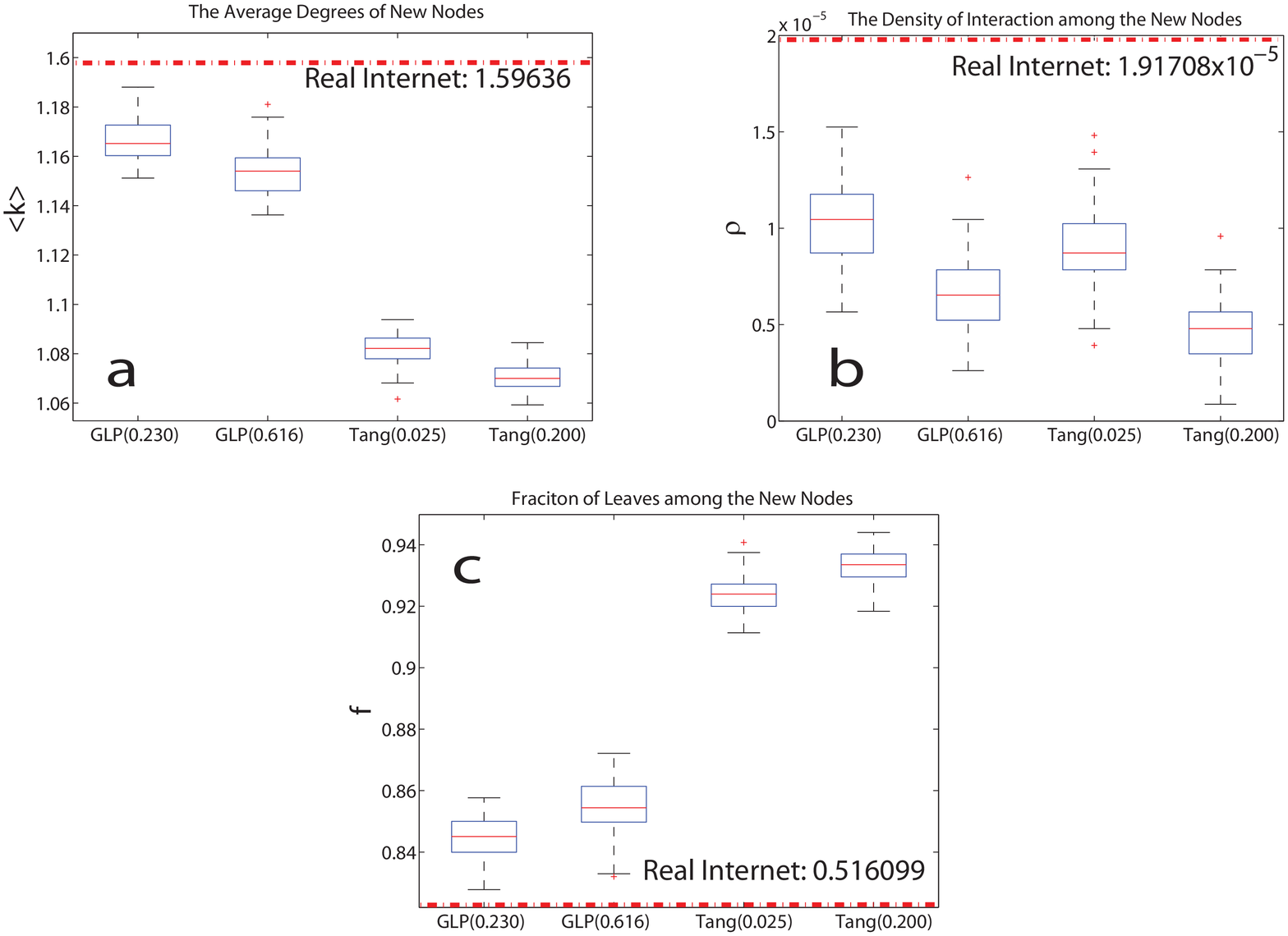}
\caption{(a) The average degree of the newly generated nodes; (b)
The density among the newly generated nodes; (c) The fraction of
leaves in the newly generated nodes. Dash line in each plot
represents the values for the real Internet. The structural
features corresponding to the networks obtained by our suggesting
parameters are closer to the reality. For each model with each
parameter, we generate $100$ networks and use the so-called
box-and-whisker plot \cite{boxplot} to display the results, where
the horizontal lines from top to bottom respectively stand for the
maximum, the upper quartile, the median, the lower quartile and
the minimum of a set of data.} \label{fig:figure2}
\end{figure*}

In this paper we focus on the models of the AS-level Internet. Two
popular models - Generalized Linear Preferential model (GLP)
\cite{Bu_infocom_2002} and Tel Aviv Network Generator (Tang)
\cite{Bar_lncs_2004} - will be evaluated by our method. The
well-known Barab\'asi-Albert (BA) \cite{Barabasi_science_1999} and
Erd\"os-R\'enyi (ER) \cite{Erdos_1960, Bollobas_1985} models are
also analyzed as two benchmarks.

The data sets we utilize here are collected by the
\emph{Routeviews Project} \cite{RP_website}. We use the data of
Jun. 2006 and Dec. 2006. Some nodes and edges in Jun. 2006
disappear in the record of Dec. 2006. Although an autonomous
system might be canceled, rarely does it happen during a short
time span. Therefore we assume that the nodes and edges in Jun.
2006 will not disappear in Dec. 2006. That is to say that the
network configuration in Jun. 2006 is a subgraph of that in Dec.
2006. We merge the network of Jun. 2006 into that of Dec. 2006 to
make a set substraction between the two sets to obtain the newly
generated edges and nodes. The basic information of the processed
data set of Dec. 2006 and two original data sets is shown in Table
\ref{tab:table1}.

\begin{table}\makeatletter\def\@captype{table}\makeatother
    \caption{The number of nodes and edges of the three data sets: two real data sets and one data set that is processed as we describe in the paper. }
    \begin{tabular}{c | c | c}
    \toprule Time & \# Nodes & \# Edges\\
    \hline
    2006.06 & 22960 & 49545\\
    2006.12 & 24403 & 52826\\
    2006.12 (processed) & 25103 & 59268\\
    \botrule
    \end{tabular}
    \label{tab:table1}
\end{table}

Now we will describe how to calculate the likelihood of each
newly-generated edge in terms of the four models. (i) \textbf{GLP
model} - This model starts from a few nodes. At each time step,
with the probability $1-p$, one new node is added and $m$ edges
are generated between the new node and $m$ old ones and with the
probability $p$, $m$ edges are generated among the existing nodes.
The ends of new edges are selected following the rule of
generalized linear preferential attachment as
\begin{equation}\label{eq:c}
\Pi_{i}=\frac{k_{i}-\beta}{\sum_{j}{(k_{j}-\beta)}},
\end{equation}
in which $\beta\in(-\infty,1)$.
In our method if the end  $i$ of a new edge is selected among the existing nodes, then $\Pi_{i}$ is calculated by the Eq. (\ref{eq:c}). Otherwise, if the end $i$ itself is a new node, $\Pi_{i}$ is $1$. So the likelihood of a new edge connecting two existing nodes $a$ and $b$ is
\begin{equation}\label{eq:d}
P_{(a,b)} = \frac{k_{a}-\beta}{\sum_{j}{(k_{j}-\beta)}} \frac{k_{b}-\beta}{\sum_{j}{(k_{j}-\beta)}}.
\end{equation}
The likelihood of an edge generated between a new node $b$ and an existing node $a$ is
\begin{equation}\label{eq:e}
P_{(a,b)}=\frac{k_{a}-\beta}{\sum_{j}{(k_{j}-\beta)}}.
\end{equation}
When a new edge connects two new nodes $a$ and $b$, its likelihood is
\begin{equation}\label{eq:f}
P_{(a,b)}=1.
\end{equation}
(ii) \textbf{Tang model} -  This model applies a super linear
preferential mechanism, say
\begin{equation}\label{eq:g}
\Pi_{i}=\frac{k^{1+\epsilon}_{i}}{\sum_{j}{k^{1+\epsilon}_{j}}}.
\end{equation}
This model also starts with a few nodes and at each time step a new node is generated with one edge connecting to one of the existing nodes that is selected with the probability described in Eq. (\ref{eq:g}). The remaining $m-1$ edges are added between the existing nodes. For these $m-1$ nodes, one end is selected according to Eq. (\ref{eq:g}), while the other one is selected randomly. Hence the likelihood of a new edge between existing nodes is
\begin{equation}\label{eq:h}
P_{(a,b)}=\frac{1}{N} \sqrt{\frac{k_{a}^{1+\epsilon}}{\sum_{j}{k_{j}^{1+\epsilon}}}\frac{k_{b}^{1+\epsilon}}{\sum_{j}{k_{j}^{1+\epsilon}}}},
\end{equation}
where $N$ is the current size of the monitored network. Eq.
(\ref{eq:h}) takes a geometric mean due to the fact that either
$a$ or $b$ could be the one selected randomly. The cases involving
new nodes are managed in the same way as that for the GLP model.
(iii) \textbf{BA model} - The BA model also starts from a small
graph and at each time step a new node associated with $m$ edges
is added. The probability that the existing node $i$ is selected
is
\begin{equation}\label{eq:i}
\Pi_{i}=\frac{k_{i}}{\sum_{j}{k_{j}}}.
\end{equation}
Note that the original BA model cannot deal with the situation where edges are generated between two existing nodes. We thus generalize the BA model as if one edge is generated between two existing nodes, one node is selected preferentially following the Eq. (\ref{eq:i}) and another one is selected randomly. Therefore the likelihood of an edge between two existing nodes $a$ and $b$ is calculated as
\begin{equation}\label{eq:j}
P_{(a,b)}=\frac{1}{N} \sqrt{\frac{k_{a}}{\sum_{j}{k_{j}}} \frac{k_{b}}{\sum_{j}{k_{j}}}}.
\end{equation}
The likelihood of an edge connecting a new node $b$ and an old one $a$ is
\begin{equation}\label{eq:k}
P_{(a,b)}=\frac{k_{a}}{\sum_{j}{k_{j}}}.
\end{equation}
The likelihood of a new edge generated between two new nodes is $1$ as discussed above.
(iv) \textbf{ER model} - The mechanism of this model is that when one edge is generated, both its ends are selected in a random fashion. The likelihood of one edge $(a,b)$ between two old nodes is
\begin{equation}\label{eq:l}
P_{(a,b)}=\frac{1}{N^{2}}.
\end{equation}
The calculation of other two types of edges is similar to that of GLP. Note that BA is a special case equivalent to the GLP model when $\beta=0$. It is also obvious that the ER model is a special case of the Tang model when $\epsilon=0$.

\begin{table}\makeatletter\def\@captype{table}\makeatother
    \caption{Maximum likelihoods and the corresponding parameters for the four models.}
    \begin{tabular}{c|c|c}
    \toprule
    Model & Maximum Likelihood & Optimum parameters\\
    \colrule
    GLP & $3.54 \times 10^{-120497}$ & $0.230$\\
    Tang & $9.77 \times 10^{-124442}$ & $0.025$\\
    ER & $4.17 \times 10^{-132356}$ & N/A\\
    BA & $2.26 \times 10^{-124449}$ & N/A\\
    \botrule
    \end{tabular}\label{tab:table2}
\end{table}

The likelihoods of the four evolving models with different
parameters are shown in Figure \ref{fig:figure1}. The maximum
likelihoods as well as the corresponding parameters are listed in
Table \ref{tab:table2}. The maximum likelihoods of both specific
Internet models (GLP and Tang) are greater than those of the BA
model and the ER model. Notice that the BA and ER model are
parameter-free and thus represented by two straight lines in
Figure \ref{fig:figure1}. Our results suggest that subject to the
mimicking of the AS-level Internet evolution, the GLP model is
better than the Tang model, and the Tang model is better than the
BA model, of course, the ER model performs the worst. A puzzling
point is that the optimal parameters corresponding to the maximum
likelihoods are far from the ones suggested in the original
literature \cite{Bu_infocom_2002, Bar_lncs_2004}. We next devise
an experiment to demonstrate that the parameters obtained by our
method are more advantageous than the original ones.

Traditionally, an evolving model starts from a small network with
a few nodes. In this experiment, we respectively use the GLP and
Tang models to drive the network evolution starting from the
configuration of Jun. 2006, ending with the same size of the
configuration of Dec. 2006. According to the Refs.
\cite{Bu_infocom_2002, Bar_lncs_2004} and the data, $\beta=0.616,
m=1.13, p=0.5214$ and $\epsilon=0.2$. Then we analyze some
statistical features of the newly generated part including the
average degree, the density of interaction and the fraction of
leaves. We find that the performance of the GLP model is better
than the Tang model with the same kind of parameters in the three
cases, demonstrating that our evaluating method is reasonable. For
both the two models, the statistical features obtained by the
optimum parameters suggested by us resemble the real data better
than those obtained by using the original parameters. The
comparisons are shown in Figure \ref{fig:figure2}.

\section{Conclusion and Discussion}

Thousands of network models are put forward in recent ten years.
Some of them aim at uncovering mechanisms that underlie general
topological properties like scale-free nature and small-world
phenomenon, others are proposed to reproduce structural features
of specific networks, such as the Internet, the World Wide Web,
co-authorship networks, food webs, protein-protein interacting
networks, metabolic networks, and so on. Besides the prosperity,
we are worrying that there is no unified method to evaluate the
performance of different models, even if the target network is
given beforehand.

Instead of considering many structural metrics, this paper reports
an evaluating method based on likelihood analysis, with an
assumption that a better model will assign a higher likelihood to
the observed structure. We have tested our method on the real
Internet at the AS level, and the results suggest that the GLP
model outperforms the Tang model, and both models are better than
the BA and ER models. This method can be further applied in
determining the optimal parameters of network models, and the
experiment indicates that the parameters obtained by our method
can better capture the structural characters of newly-added nodes
and links.

The main contributions of this work are twofold. In the
methodology aspect, we provide a starting point towards a unified
way to evaluate network models. In the perspective aspect, we
believe for majority of real evolutionary networks, the driven
factors and the parameters will vary in time. For example, recent
empirical analysis suggests that before and after the year 2004,
the Internet at the AS level grows with different mechanisms
\cite{Zhang_njp_2008}. To find out a single mechanisms that drives
a network from a little baby to a giant may be an infeasible task.
In fact, in different stages, a network could grow in different
ways, or in a hybrid matter with changing weight distribution on
several mechanisms. Once, the research focus has shifted from
analyzing static models to evolutionary models. In the near
future, it may shift from the evolutionary models to the evolving
of the evolutionary models themselves. In principle, the current
method could capture the tracks of not only the network evolution,
but also the mechanism evolution. Hopefully this work could
provide some insights into the studies on network modeling.

\begin{acknowledgments}
We acknowledge A. Wool for the codes of the Tang model. This work
is supported by the National Natural Science Foundation of China
under grant No. 11075031 and the Fundamental Research Funds for
the Central Universities
\end{acknowledgments}

\end{document}